The essential signature of a massive starburst in a distant quasar

P. Solomon*, P. Vanden Bout†, C. Carilli‡, and M. Guelin#

*Department of Physics and Astronomy, SUNY at Stony Brook, Stony Brook NY 11794, USA; †National Radio Astronomy Observatory, 520 Edgemont Rd., Charlottesville VA 22903, USA; ‡National Radio Astronomy Observatory, P.O. Box 0, Socorro NM 87801, USA; #Institut de Radioastronomie Millimétrique, 300 rue de la Piscine, Domaine Universitaire, 38406 Saint Martin d'Hères, France.

**Observations of carbon monoxide (CO) emission in high redshift ($z>2$) galaxies indicate the presence of large amounts of molecular gas. Many of these galaxies contain an active galactic nucleus (AGN) powered by accretion of gas onto a supermassive black hole, and a key question is whether their extremely high infrared luminosities result from the AGN, or from bursts of massive star formation (associated with the molecular gas), or both. In the Milky Way, high-mass stars form in the dense cores of interstellar molecular clouds; gas densities are $n(H_2)>10^5$ cm$^{-3}$ in the cores[1,2]. Recent surveys show that virtually all galactic sites of high-mass star formation have similarly high densities[3]. The bulk of the cloud material traced by CO observations is at a much lower density. In galaxies in the local Universe, the HCN(J=1-0) line is an effective tracer of the high-density molecular gas.[4] Here we report observations of HCN emission in the early Universe from the infrared luminous 'Cloverleaf' quasar (at a redshift $z=2.5579$). The HCN line luminosity indicates the presence of 10 billion solar masses of very dense gas, an essential**

**feature of an immense starburst that contributes, together with the AGN it harbors, to its high infrared luminosity.**

The Cloverleaf is one of about 25 galaxies observed in the epoch of galaxy formation (z>2) in CO emission. The Cloverleaf (also known as H1413+117) has its optical and CO emission imaged into four components by a gravitational lens. The first detection of the Cloverleaf in molecular line emission was that of CO(J=3-2) by Barvainis et al.[5] It has now been observed[6] in four transitions of CO, with a weighted mean redshift for the CO emission of z=2.5579, and in the atomic carbon fine structure lines[6,7] as well. A recently constructed source-lens model (see ref. 8 for a discussion of this and other Cloverleaf lens models), based on high resolution images of CO(J=7-6) emission, shows that the Cloverleaf contains an extensive molecular disk with a diameter of 1.6 kpc.

CO emission, in this source and in general, is the best tracer of molecular clouds, the potential fuel for star formation and a major component of interstellar matter. But the low density required for excitation of CO, $n(H_2) > 300$ cm$^{-3}$, means it is not a specific tracer of molecular cloud cores or star formation rates. In contrast, HCN is a specific tracer of cloud cores. The first observed correlation[4] between HCN(J=1-0) luminosity and infrared luminosity (IR) for galaxies has recently been expanded to cover many more sources[9]. The HCN-IR correlation for galaxies, which extends over three orders of magnitude in $L_{IR}$, is linear. Unlike $L_{IR}/L'_{CO}$, the ratio of $L_{IR}/L'_{HCN}$ does not increase with the IR luminosity, indicating that the star formation rate per unit dense molecular gas is the same for ultra-luminous galaxies and normal spirals. Observation of a high HCN line luminosity would strongly indicate the presence of high-mass star formation in the Cloverleaf, and to that end we have used the NRAO Very Large Array (VLA) to search for HCN(J=1-0) emission.

The VLA image (Fig. 1) of line plus continuum for 7 spectral channels shows the line clearly at the position of the quasar in the 4 center channels. The spectrum of the HCN emission (Fig. 2) has a similar profile to that of the CO(J=7-6) emission.[10] The HCN emission is clearly detected with a peak line flux density six times the rms noise, S(peak)=0.24±0.04 mJy.

Table 1 summarizes the observed and derived (intrinsic) quantities, including the ratios $L_{IR}/L'_{HCN}$ and $L'_{HCN}/L'_{CO}$. The line luminosities L′ are a measure of the surface brightness times area in brightness temperature units[11]. Two lines with the same spatial extent, velocity profile, and brightness temperature $T_b$ will have the same L′ regardless of transition or line frequency. All the luminosities in Table 1 are corrected for lens magnification, using the results of a lens model[8] based directly on a CO(J=7-6) image. This model leads to a derived CO disk diameter of 1.6 kpc, similar to that of the CO emitting regions in nearby starburst ultra-luminous galaxies[12]. It should be noted, that the diameter of the dust emission region, responsible for producing the infrared luminosity, has been determined[13] to be 1.5 kpc from a comparison of models of radiation transfer in the dust with the observed spectral energy distribution. Thus, both the molecular and infrared emission originate in extended regions of the same size. In this respect, the

Cloverleaf is similar to PSS J2322+1944, where the molecular gas in a ring surrounding a QSO has been directly imaged using gravitational lensing[14].

The Cloverleaf HCN(1-0) line luminosity is larger by a factor of 100 than that of normal spiral galaxies. The only galaxies with HCN line luminosities comparable to that of the Cloverleaf are ultra-luminous infrared galaxies. The intrinsic HCN line luminosity of the Cloverleaf is about a factor of two higher than that of the low redshift ultra-luminous IR galaxy Mrk 231, the most luminous object in the local universe, and a factor of three higher than that of Arp 220, the prototype ultra-luminous IR galaxy. To put this in perspective, the Cloverleaf intrinsic HCN line luminosity is ten times greater than the CO line luminosity of the Milky Way. A large HCN line luminosity is a clear indicator of a large mass of dense star-forming molecular gas. Assuming that the HCN emission originates in dense gravitationally bound regions of average density $n(H_2)=3\times10^4$ cm$^{-3}$ and intrinsic brightness temperature $T_b=50K$, the mass of dense gas is about 7 times $L'_{HCN}$, or $2\times10^{10}\,M_\odot$. (The gas mass scales as $n^{1/2}/T_b$ for gravitationally bound clouds.)

In star formation models, the star formation rate is proportional to the infrared luminosity of dust heated by embedded recently formed high-mass stars. The ratio $L_{FIR}/L'_{HCN}$ is a measure of the star formation rate per unit mass of dense gas.[4] $L'_{CO}$ is an indicator of the total molecular gas, because the CO transitions are so easily excited. The ratio $L'_{HCN}/L'_{CO}$ is a measure of the fraction of all molecular gas that is in dense star-forming cloud cores. This fraction is much higher in ultra-luminous IR galaxies than in normal spirals. For the Cloverleaf, $L_{IR}/L'_{HCN}$ is about a factor of 5 times higher than typical ultra-luminous IR galaxies and $L'_{HCN}/L'_{CO}$ is half that of Arp 220 and a factor three less than that of Mrk 231. In making this comparison we have assumed the brightness temperature of the CO(J=1-0) line is 1.1 times the J=3-2 line.[7] (One possibility that we have not considered is that the magnification for the molecular emission and the IR emission are different, which could distort the intrinsic ratios. The similar sizes deduced for the molecular disk and the dust emission is evidence that the magnification is the same for both.)

The molecular and IR luminosities for the Cloverleaf show that star formation in the large mass of dense molecular gas indicated by the HCN luminosity could account for a substantial fraction, but not all, of the IR luminosity from this quasar. Using Arp 220 as a standard for the luminosity ratio $L_{FIR}/L'_{HCN}$, star formation in the dense molecular gas could account for $5 \times 10^{12}$ $L_\odot$ or about 20% of the total intrinsic IR luminosity. Using normal spirals as a standard for the ratio[4,9] $L_{IR}/L'_{HCN}$ would lead to a lower limit of 10%, while using the highest ratio for an ultra-luminous IR galaxy gives an upper limit of 40%. A recent model[7] of the IR spectral energy distribution of the Cloverleaf has two distinct components: one with a warm dust temperature $T_d=115K$ responsible for the mid-IR, and a second much more massive component with $T_d=50K$ that produces the far-IR. The far-IR luminosity, 22% of the total, may correspond to the luminosity generated by star formation deduced above. The ratio $L_{FIR}/L'_{HCN}=1700$ is then comparable to that of ultra-luminous IR galaxies.

Adopting the Arp 220 comparison for the expected ratio $L_{FIR}/L'_{HCN}$ yields a star formation rate of $10^3$ $M_\odot$ per year for the Cloverleaf, 300 times the rate in the Milky Way and 30 times greater than that of optical-UV starburst galaxies. The dense gas depletion time or starburst lifetime is about $10^7$ years. Thus, in addition to the quasar, the

Cloverleaf galaxy appears to have a huge starburst, more luminous than any optical starburst and comparable to that in ultra-luminous IR galaxies.

___

**Figure 1** HCN(J=1-0) channel maps from the Cloverleaf observations obtained with the NRAO Very Large Array. Contour levels are –3, -2, -1, 1, 2, 3, 4, 5, 6, 7, 8 x 0.07 mJy/beam (equivalent to 1.2 times the rms noise, except for channel 7 which has an rms noise of 0.09 mJy). The continuum has not been subtracted. The naturally weighted beam has a FWHM = 4 arc-seconds. The channel width is 6.25 MHz, equivalent to 75 Km s$^{-1}$; the channels are centered at +231, +154, … -231 km/s with respect to an observed frequency of 24.9149 GHz. The coordinates of the position marked by a cross are: R.A. 14:15:46.27, Dec. +11:29:43.5; J2000. The observations were made in the D-configuration of the VLA on March 10, 16, and 31 (2003), with total integration time of 12 hours.

**Figure 2** Spectrum of HCN(J=1-0) emission observed in the Cloverleaf together with the observed[9] spectrum of CO(J=7-6), scaled down by a factor of 200. Zero velocity corresponds to an observing frequency of 24.9149 GHz. The general similarity of the two line profiles, including the negative velocity excess emission seen most clearly in the CO line, suggests common kinematics for the emission regions of the HCN line and this high excitation CO line. A continuum of 0.26 mJy has been subtracted. The continuum was determined by an observation at 22.5 GHz and extrapolated to 24 GHz using the spectral index of –1.35 measured between 8 and 22 GHz. The result is $S_{CONT}$(24GHz) =

0.26±0.03 mJy, where the error includes both the errors in the measurement at 22.5 GHz and the spectral index.

ACKNOWLEDGEMENTS: The National Radio Astronomy Observatory is operated by Associated Universities, Inc., under cooperative agreement with the National Science Foundation. P. V. B. thanks Columbia U. and the Institut d'Astrophysique de Paris for hospitality during this research. The authors thank D. Downes for useful discussions.

COMPETING INTEREST STATEMENT: The authors declare that they have no competing financial interests.

CORRESPONDENCE should be addressed to:

P. M. S. (e-mail: psolomon@sbastk.ess.sunysb.edu)

**Table 1 – Observed and Derived (Intrinsic) Quantities**

| Observed Quantities: HCN (1-0) | H 1413+117 | Mrk 231 | Arp 220 |
|---|---|---|---|
| z (HCN)[a] | 2.5569±0.0006 | | |
| S (HCN, peak)[b], mJy | 0.24±0.04 | | |
| I(HCN) = S ()v), Jy km s$^{-1}$ | 0.069±0.012 | | |
| Derived[c] Intrinsic Quantities: | | | |
| L´(HCN, J=1-0), K km s$^{-1}$ pc$^2$ | (3.2±0.5) x 10$^9$ | 2 x 10$^9$ | 1.1 x 10$^9$ |
| L(IR)[c]/ L´(HCN), L$_\odot$ / K km s$^{-1}$ pc$^2$ | 7700±1300 | 1600 | 1500 |
| L(FIR)[c]/ L´(HCN), L$_\odot$ / K km s$^{-1}$ pc$^2$ | 1700±300 | 1100 | 1300 |
| L´(HCN)/L´(CO)[d] | 0.07±0.01 | 0.24 | 0.12 |

[a] Taking line center halfway between two peak channels; rest frequency of 88.632 GHz.

[b] A tentative detection of the HCN(J=4-3) line has been reported (ref. 6) with a peak line flux density of S = 5.5±1.1 mJy, based on observations made with the IRAM 30m Telescope. Our own IRAM Interferometer observations of HCN(J=4-3) show no clear indication of a line. The interferometer data yields an upper limit of 3 mJy (3 x rms noise) for the S(HCN 4-3, peak).

[c] Throughout this paper we have used a luminosity distance of 21 Gpc (flat universe with H$_0$=70, $\Sigma_m$= 0.3 and $\Sigma_7$= 0.7) and a lens magnification of 11 (ref. 8).

Errors quoted for the derived quantities reflect the error in the precision of the integrated flux I(HCN). There is also a possible systematic error from subtraction of the continuum with an uncertainty (2rms) of 0.06 mJy. This would affect the accuracy of the integrated quantities by 25 %. (see fig 1 and 2)

[d] Total infrared (10-2000:m), far-infrared (42.5-122.5:m), and CO(J=1-0) luminosities for the Cloverleaf (rest frame wavelengths) from data in ref. 7; the IR/FIR and HCN luminosities in Mrk231 and Arp220 from refs. 9 and 4 respectively.

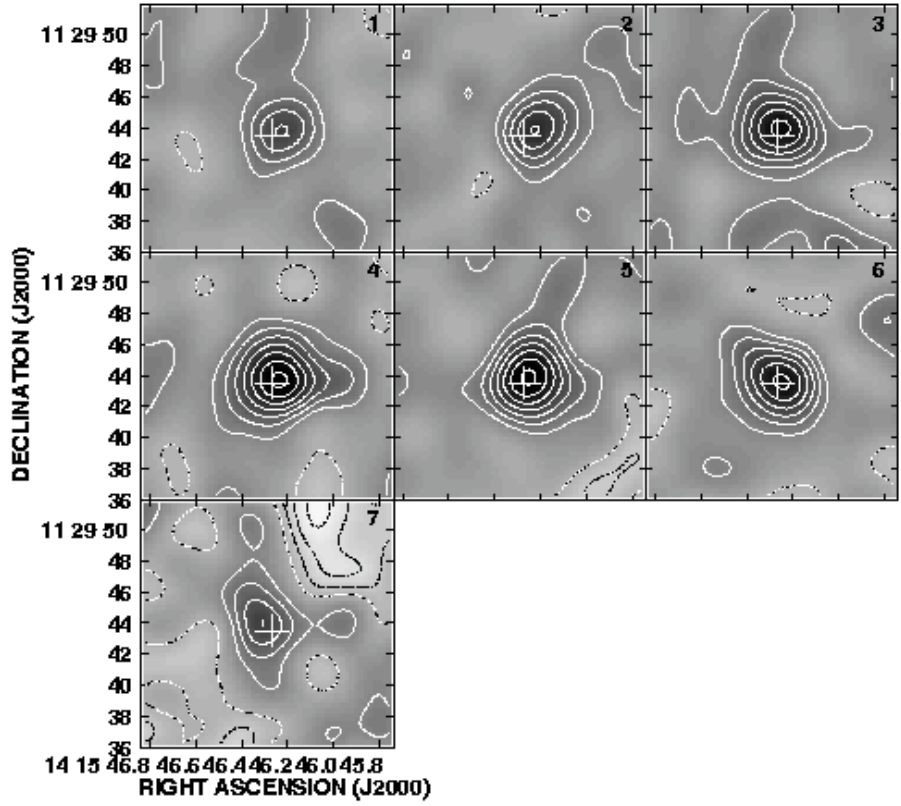

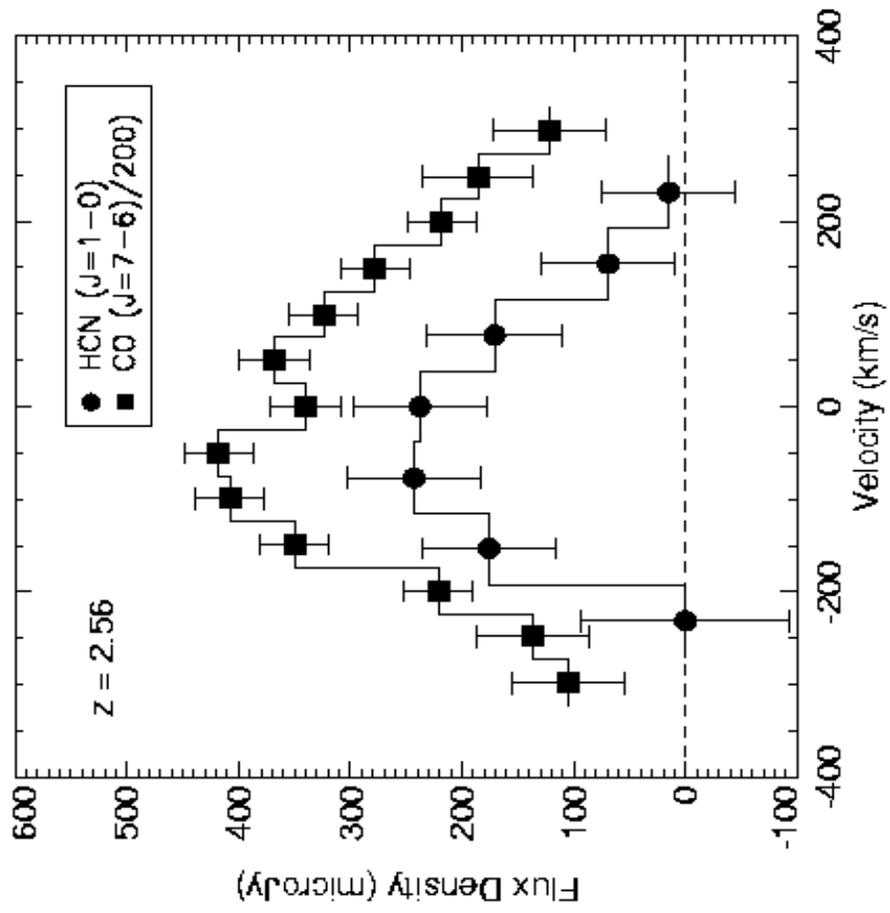